\documentclass[10pt,twocolumn,showpacs,amsmath,amssymb,floatfix,superscriptaddress]{revtex4-2}

\usepackage{graphicx}
\usepackage{dcolumn}
\usepackage{bm}
\usepackage{color}
\usepackage{txfonts}
\usepackage{microtype}
\usepackage{hyperref}
\usepackage[english]{babel}
\usepackage{slashed}
\usepackage{gensymb}
\usepackage{epsfig}
\usepackage[normalem]{ulem}
\usepackage{amsmath}
\usepackage{array}
\usepackage{booktabs}
\allowdisplaybreaks[4]

\begin{document}
\renewcommand{\figurename}{FIG}	

\title{Cascade of polarized Compton scattering and Breit-Wheeler pair production}

\author{Qian Zhao}
\affiliation{Ministry of Education Key Laboratory for Nonequilibrium Synthesis and Modulation of Condensed Matter, Shaanxi Province Key Laboratory of Quantum Information and Quantum Optoelectronic Devices, School of Physics, Xi'an Jiaotong University, Xi'an 710049, China}
\author{Ting Sun}
\affiliation{Ministry of Education Key Laboratory for Nonequilibrium Synthesis and Modulation of Condensed Matter, Shaanxi Province Key Laboratory of Quantum Information and Quantum Optoelectronic Devices, School of Physics, Xi'an Jiaotong University, Xi'an 710049, China}
\author{Kun Xue}
\affiliation{Ministry of Education Key Laboratory for Nonequilibrium Synthesis and Modulation of Condensed Matter, Shaanxi Province Key Laboratory of Quantum Information and Quantum Optoelectronic Devices, School of Physics, Xi'an Jiaotong University, Xi'an 710049, China}
\author{Feng Wan}\email{wanfeng@xjtu.edu.cn}
\affiliation{Ministry of Education Key Laboratory for Nonequilibrium Synthesis and Modulation of Condensed Matter, Shaanxi Province Key Laboratory of Quantum Information and Quantum Optoelectronic Devices, School of Physics, Xi'an Jiaotong University, Xi'an 710049, China}
\author{Jian-Xing Li}\email{jianxing@xjtu.edu.cn}
\affiliation{Ministry of Education Key Laboratory for Nonequilibrium Synthesis and Modulation of Condensed Matter, Shaanxi Province Key Laboratory of Quantum Information and Quantum Optoelectronic Devices, School of Physics, Xi'an Jiaotong University, Xi'an 710049, China}

\date{\today}
	
\begin{abstract}
Cascaded Compton scattering and Breit-Wheeler (BW) processes play fundamental roles in high-energy astrophysical sources and laser-driven quantum electrodynamics (QED) plasmas. A thorough comprehension of the polarization transfer in these cascaded processes is essential for elucidating the polarization mechanism of high-energy cosmic gamma rays and laser-driven QED plasmas. In this study, we employ analytical cross-sectional calculations and Monte Carlo (MC) numerical simulations to investigate the polarization transfer in the cascade of electron-seeded inverse Compton scattering (ICS) and BW process. Theoretical analysis indicates that the polarization of background photons can effectively transfer to final-state particles in the first-generation cascade due to helicity transfer. Through MC simulations involving polarized background photons and non-polarized seed electrons, we reveal the characteristic polarization curves as a function of particle energy produced by the cascaded processes of ICS and BW pair production. Our results demonstrate that the first-generation photons from ICS exhibit the non-decayed stair-shape polarization curves, in contrast to the linearly decayed ones of the first-generation electrons. Interestingly, this polarization curve trend can be reversed in the second-generation cascade, facilitated by the presence of polarized first-generation BW pairs with fluctuant polarization curves. The cascade culminates with the production of second-generation BW pairs, due to diminished energy of second-generation photons below the threshold of BW process. Our findings provide crucial insights into the cascaded processes of Compton scattering and BW process, significantly contributing to the understanding and further exploration of laser-driven QED plasma creation in laboratory settings and high-energy astrophysics research.
\end{abstract}
	\maketitle	
\section{Introduction}\label{introduction}
Electron-positron ($e^+e^-$) pair plasmas play a crucial role in the realm of high-energy astrophysics, particularly in the most extreme celestial sources such as active galactic nuclei, gamma-ray bursts (GRBs), black hole binaries, and pulsars \cite{Ruffini2010,Kumar2015,Kostenko2018,Lundman2018}. Moreover, the creation of pair plasmas holds significant potential in the forthcoming generation of tens-petawatt lasers, which boast intensities surpassing $10^{23}$ W/cm$^{2}$ \cite{Zhang2020,Yoon2021}. The formation of pair plasmas hinges upon the cascaded interactions of Compton scattering and Breit-Wheeler (BW) pair production, which are recognized as the two fundamental mechanisms \cite{Hirotani1998,Chen2018,Chen2020,Crinquand2020,Galishnikova2023}. For instance, within the black hole magnetosphere, relativistic leptons undergo acceleration within unscreened electrostatic gaps generated by the black hole's rotation, attaining relativistic energies. Subsequently, the relativistic leptons engage in inverse Compton scattering (ICS) with soft background photons emitted by the accretion disk. This process yields high-energy gamma-ray photons, which, in turn, collide with other background photons, resulting in the production of $e^+e^-$ pairs. These cascading processes continually replenish the pair plasma, thereby sustaining the force-free magnetosphere, which serves as a model for plasma jet launching in black holes.

Relativistic jets encompass various astrophysical phenomena, with GRBs being one notable example. Accurate measurement of the polarization in GRBs is crucial for attaining a comprehensive understanding of the physical characteristics of these jets, including the energy dissipation locations and radiation mechanisms \cite{Mignani2019,Mao2017,Lan2019}. High levels of linear polarization have been observed in the prompt emissions of GRBs at high-energy bands. For instance, RHESSI observations have reported polarization levels of up to $80\%$ \cite{Coburn2003}, while INTEGRAL observations have detected polarization of $98\%$ within the energy range of 100-350 keV \cite{Kalemci2007}. Furthermore, circular polarization of approximately $60\%$ has been observed in the optical afterglow \cite{Wiersema2014}. The polarization in GRBs is commonly interpreted through synchrotron emission, which can arise from various magnetic field configurations within structured GRB jets, including aligned, toroidal, and random magnetic fields \cite{Nava2016}. Notably, polarized synchrotron soft photons can undergo ICS by cold relativistic electrons within the outflow, resulting in the production of hard gamma-ray photons \cite{Chang2013}.

The advent of petawatt laser facilities around the world \cite{ELI,ELI_np,Apollon,Yoon2021} has opened the door to ultra-relativistic laser-plasma interactions in the regime dominated by strong-field quantum electrodynamics (QED). In this regime, nonlinear QED processes can give rise to the production of abundant gamma-ray photons and electron-positron ($e^+e^-$) pairs \cite{Piazza2012,Fedotov2023}. The electromagnetic QED cascade, characterized by the emission of multiple photons and their subsequent conversion into $e^+e^-$ pairs, has been extensively investigated in vacuum, with a particular focus on laser intensities approaching the critical field strength for vacuum pair production \cite{Fedotov2010,Elkina2011,Bulanov2013,Tamburini2017}. Accounting for the polarization of photons and $e^+e^-$ pairs, the electromagnetic QED cascade in vacuum can produce highly polarized particles and induce asymmetries in the polarization distribution \cite{King2013,Seipt2021}. The coupling between relativistic collective plasma dynamics and electromagnetic QED cascade processes can lead to the creation of dense pair plasmas \cite{Nerush2011,Ridgers2012,Sarri2015,Luo2015,Jirka2017,Luo2018,Gu2019}, enabling laboratory investigations of high-energy astrophysical phenomena, such as GRBs, using upcoming tens-petawatt lasers. Recent studies have shown considerable interest in polarized strong-field QED processes in laser-driven plasma \cite{Sun2022,Gonoskov2022,Fedotov2023}, which have been incorporated into particle-in-cell (PIC) simulations to explore polarization effects within strong electromagnetic plasma environments. Moreover, QED-PIC simulations indicate that irradiating solid targets with petawatt lasers at peak intensities exceeding $10^{23}$ W/cm$^{2}$ can lead to the production of dense polarized gamma-ray photons and polarized $e^+e^-$ pairs \cite{Xue2020,Song2021,Song2022,Xue2023,Wan2023}.

The production of laser-driven QED plasmas, characterized by high-density polarized gamma-ray photons and $e^+e^-$ pairs exceeding $10^{16}$/ cm$^{-3}$ \cite{Sarri2015}, highlights the significance of binary QED collisions through the cascade of Compton scattering and BW pair production in the subsequent dynamic evolution of pair plasma. Preliminary investigations employing 3D PIC simulations suggest that the linear BW process could dominate pair production during the creation of laser-driven QED plasmas \cite{He2021phys,He2021njp}. Hence, in order to attain a comprehensive understanding of the polarization physics in GRBs and perform accurate simulations of laser-driven pair plasmas, it is crucial to establish complete knowledge of the cascade between polarized Compton scattering and BW process. The complete polarization effects of the BW process have been investigated using fully angle- and spin-resolved Monte Carlo (MC) numerical methods \cite{Zhao2022,Zhao2023}. To model the cascade between polarized Compton scattering and BW process, the former can be described utilizing the same numerical method with fully polarized cross sections \cite{Berestetskii1982}. Prior to undertaking fully consistent simulations with the QED-PIC method \cite{Wan2023}, it is necessary to carefully examine the polarization-associated physics of cascaded processes of Compton scattering and BW pair production within a charge-free space.

In this paper, we investigate the cascaded processes of polarized Compton scattering and BW pair production using analytical cross sections and MC numerical simulations. The polarization transfer in the cascade of multiple Compton scattering and BW process is examined in the center-of-momentum (c.m.) frame through theoretical calculations. These calculations reveal that the polarization transfer can be enhanced in the cascaded processes through the helicity transfer of photons and electrons. We visualize the electron-seeded cascade of ICS and BW process using multi-parameter MC simulations, assuming non-polarized seed electrons and background soft photons with different polarization. The resulting first-generation gamma-ray photons from multiple ICS exhibit non-decayed stair-shape polarization curves with respect to photon energy, in contrast to the linearly decayed polarization of first-generation electrons. The first-generation BW pairs, produced from collisions between first-generation photons and background photons, possess energy-dependent fluctuant polarization curves. The characteristic polarization and spectrum of the produced BW pairs, upon colliding with background photons, give rise to an exponential spectrum of second-generation photons with linearly decayed polarization curves. Additionally, the resulting second-generation electrons exhibit saturated polarization curves. This comprehensive understanding of the cascaded processes of Compton scattering and BW pair production is highly beneficial for investigating upcoming laser-driven QED plasmas in the laboratory and for associated research in high-energy astrophysics.

The paper is organized as follows: In Sec.~\ref{theory}, we analytically investigate the polarization transfer in the cascaded processes of Compton scattering and BW pair production using completely polarized cross sections. In Sec.~\ref{simulation}, we employ MC simulations to explore the energy-dependent polarization of the produced gamma-ray photons and BW pairs in the cascade of multiple ICS and BW process. Finally, we present a concise summary of our findings in Sec.~\ref{summary}.
\section{Theoretical analysis of polarization transfer in the cascaded processes}\label{theory}
\begin{figure}[!t]
	\setlength{\abovecaptionskip}{0.2 cm}
	\centering\includegraphics[width=1\linewidth]{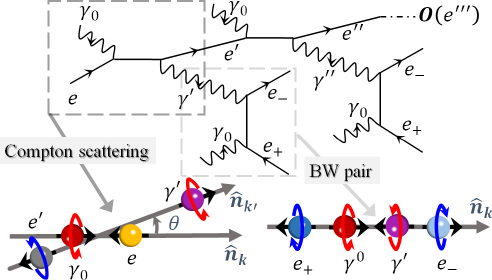}
	\caption{
Top Feynman diagrams: cascaded processes in which the scattered gamma-ray photons ($\gamma'$,$\gamma''...$) from multiple Compton scattering collide with background soft photons $\gamma_0$ to produce BW $e^+e^-$ pairs. Bottom sketches: helicity transfer of the above Compton scattering and BW processes in the c.m. frame.
}
	\label{fig:sketch}
\end{figure}
Considering the collisions between a relativistic, non-polarized electron ($e$) and circularly polarized background soft photons ($\gamma_0$), multiple ICS events result in the polarization of scattered gamma photons ($\gamma', \gamma'', \ldots$) and electrons ($e', e'', \ldots$), while subsequent BW processes between $\gamma', \gamma'', \ldots$ and $\gamma_0$ lead to the production of polarized $e^+e^-$ pairs, as illustrated in Fig. \ref{fig:sketch}. Specifically, due to helicity transfer, $\gamma_0$ gives rise to the production of first-generation polarized $\gamma', \gamma'', \ldots$ and spin-polarized $e', e'', \ldots$ through multiple ICS interactions. These polarized $\gamma', \gamma'', \ldots$ then undergo collisions with $\gamma_0$, resulting in the production of first-generation $e^+e^-$ pairs via BW process. The polarized pairs subsequently Compton scatter $\gamma_0$, leading to the production of second-generation Compton photons and electrons for the subsequent cascade. In each single Compton scattering (BW process), the scattering angle $\theta$ is defined as the angle between momentum $\bm{k}$ of the initial-state  photon and momentum $\bm{k}'$ ($\bm{p}'_e$) of the final-state photon $\gamma'$ (final-state electron $e'$) in the c.m. frame.

In order to theoretically model this cascaded processes, the cross sections of Compton scattering and BW process with arbitrary polarization should be formulated.
The cross section for the BW process, accounting for arbitrary initial photon polarization and final electron (positron) spins, has been developed utilizing photon and lepton density matrices \cite{Zhao2022,Zhao2023}. Similarly, employing analogous theoretical techniques, the cross section for polarized Compton scattering, considering arbitrary polarization of initial and final particles, can be obtained (see Appendix \ref{appA} for details). In addition to the theoretical analysis, the availability of cross sections for arbitrarily polarized Compton scattering and BW process is advantageous for modeling realistic binary QED collisions using MC sampling. In the linear processes, polarization transfer is accomplished only via the helicity transfer, which is different from the strongly nonlinear QED processes driven by electron-laser interactions, where polarization transfer involves both the field structure and particle spin \cite{Li2019,Wan2020,Li2020,Xue2022,Chen2022,Zhuang2023}. Furthermore, because the angular modulations originate from the helicity transfer in the binary scattering system fundamentally, the angle dependence is nontrivial in the production of distributed polarization. However, the angle dependence in the cross sections of strong-field QED processes within framework of locally constant cross-field approximation  is always negligible because of the collimated radiation cone angle for ultra-relativistic electrons \cite{Chen2022}.

With the formulated cross sections in c.m. frame, the polarization transfer in the cascaded processes can be analytically investigated. In electron-seeded multiple Compton scattering, the polarization state of incident photon, namely, background photon $\gamma_0$, is fixed and described by Stokes parameters $(\xi^{(0)}_1,\xi^{(0)}_2,\xi^{(0)}_3)$. While the polarization of incident electron, described by spin vector $(\zeta_1,\zeta_2,\zeta_3)$, transits in each scattering.
From the spin-dependent cross section of Compton scattering [see Eq. (\ref{dsig})], one obtains the Stokes parameters of final-state photon
\begin{subequations}\label{xif}
\begin{eqnarray}
\xi_{\gamma,1}&=&\frac{1}{F_{\rm{c}}}G^\gamma_1(\zeta_1,\xi^{(0)}_2),\\
\xi_{\gamma,2}&=&\frac{1}{F_{\rm{c}}}G^\gamma_2(\zeta_1,\zeta_2,\zeta_3,\xi^{(0)}_1,\xi^{(0)}_2,\xi^{(0)}_3),\\
\xi_{\gamma,3}&=&\frac{1}{F_{\rm{c}}}G^\gamma_3(\zeta_2,\zeta_3,\xi^{(0)}_2,\xi^{(0)}_3),
\end{eqnarray}
\end{subequations}
and spin components of final-state electron
\begin{subequations}\label{zef}
\begin{eqnarray}
\zeta_{e,1}&=&\frac{1}{F_{\rm{c}}}G^e_1(\zeta_1,\zeta_2,\xi^{(0)}_1,\xi^{(0)}_3),\\
\zeta_{e,2}&=&\frac{1}{F_{\rm{c}}}G^e_2(\zeta_1,\zeta_2,\zeta_3,\xi^{(0)}_1,\xi^{(0)}_2,\xi^{(0)}_3),\\
\zeta_{e,3}&=&\frac{1}{F_{\rm{c}}}G^e_3(\zeta_1,\zeta_2,\zeta_3,\xi^{(0)}_1,\xi^{(0)}_2,\xi^{(0)}_3).
\end{eqnarray}
\end{subequations}
Here, $F_{\rm{c}}$ represents the cross section with summation over the final electron spin. The coefficients $G_i^e$ and $G_i^\gamma$ determine the transition probabilities of the polarization of the final-state electron and electron in MC sampling. The explicit expressions for these coefficients can be found in the Appendix. Because of the multiple Compton scattering,
It is important to note that, in c.m. frame, the quantities $\xi_{\gamma,2}$ and $\zeta_{e,2}$ represent the mean helicity of the photon and electron, respectively. On the other hand, $(\xi_{\gamma,1},\xi_{\gamma,3})$ and $(\zeta_{e,1},\zeta_{e,3})$ correspond to the linear polarization of the photon and the transverse spin polarization of the electron. The Stokes parameters and spin vector of the incident photon and electron share the same interpretation and significance.

With the Stokes parameters of $\gamma_0$ and scattered photon, the spin components of the final electron in the BW process can be expressed as
\begin{subequations}\label{zpf}
\begin{eqnarray}
\zeta_{-,1}&=&\frac{1}{F_{\rm{bw}}}G^-_1(\xi^{(0)}_1,\xi^{(0)}_2,\xi_{\gamma,1},\xi_{\gamma,2}),\\
\zeta_{-,2}&=&\frac{1}{F_{\rm{bw}}}G^-_2(\xi^{(0)}_2,\xi^{(0)}_3,\xi_{\gamma,2},\xi_{\gamma,3}),\\
\zeta_{-,3}&=&\frac{1}{F_{\rm{bw}}}G^-_3(\xi^{(0)}_2,\xi^{(0)}_3,\xi_{\gamma,2},\xi_{\gamma,3}),
\end{eqnarray}
\end{subequations}
where the explicit expressions for the coefficients $G_i^-$ can be found in Eq. (\ref{Gpair}). Note that the polarization of BW positron is the same to BW electron and thus only BW electron is used in analysis below.

Finally, the total polarization can be analytically calculated using the following equations:
\begin{eqnarray}
\xi_\gamma&=&\sqrt{(\xi_{\gamma,1})^2+(\xi_{\gamma,2})^2+(\xi_{\gamma,3})^2},\label{pol-gam}\\
\zeta_e&=&\sqrt{(\zeta_{e,1})^2+(\zeta_{e,2})^2+(\zeta_{e,3})^2},\label{pol-e}\\
\zeta_-&=&\sqrt{(\zeta_{-,1})^2+(\zeta_{-,2})^2+(\zeta_{-,3})^2},\label{pol-pair}
\end{eqnarray}
where $\xi_\gamma$ corresponds to the total polarization of the scattered Compton photons, $\zeta_e$ represents the total polarization of the scattered electrons, and $\zeta_-$ denotes the total polarization of the BW electrons.

\begin{figure}[!t]
	\setlength{\abovecaptionskip}{0.2cm}
	\centering\includegraphics[width=0.9\linewidth]{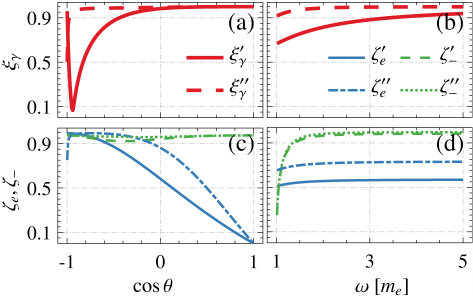}	
	\caption{Variations of final-state photon polarization and final-state lepton polarization with: (a) and (c) scattering angle $\cos{\theta}$, (b) and (d) c.m. energy $\omega$.
 $\xi'_\gamma$ and $\zeta'_e$ correspond to scattered photon $\gamma'$ and electron $e'$, and $\xi'_\gamma$ and $\zeta''_e$ correspond to $\gamma''$ and $e''$ in Fig. \ref{fig:sketch}. $\zeta_-{'}$ and $\zeta''_-$ correspond to BW electrons from $\xi'_\gamma$ and $\xi''_\gamma$ as polarization of a colliding photon. The results are obtained from Eqs. (\ref{pol-gam})-(\ref{pol-pair}) with $\xi^{(0)}_2=1$ of background photon $\gamma_0$.}
	\label{fig:cp}
\end{figure}
With the analytical polarization equations of Eqs. (\ref{pol-gam})-(\ref{pol-pair}), the polarization transfer in the cascaded processes
 can be illustrated in c.m. frame with $\omega$ of photon energy and $\cos{\theta}$ of scattering angle. For the production of first-generation photons and electron from multiple ICS of seed electrons, the final-state spin of electron in the previous scattering is used to calculate the one in the next scattering, e.g., $\zeta'_{e}$ of $e'$ from Eq. (\ref{pol-e}) is used to calculate $\xi''_\gamma$ of $\gamma''$ by Eq. (\ref{pol-gam}). Stokes parameters of final-state gamma photon in each ICS are used to calculate the spins of first-generation BW pairs by Eq. (\ref{pol-pair}), e.g., $\xi'_\gamma$ and $\xi''_\gamma$ are used to calculate $\zeta'_-$ and $\zeta''_-$ of BW electrons.

 Considering $\gamma_0$ with $\xi^{(0)}_2=1$ and non-polarized electron $e$, the polarization transfer in the first-generation cascade are shown in Fig. \ref{fig:cp}. According to Eqs. (\ref{Ggam}) and (\ref{Gele}), the produced $\gamma'$ with $\xi'_\gamma$ and $e'$ with $\zeta'_e$ can be polarized merely through helicity transfer of photon via single $\xi_2$-associated terms in $G_2^\gamma$ and $G_2^e$. As a result, $\gamma'$ and $e'$ are polarized with remarkably $\cos{\theta}$-dependent circular polarization and longitudinal polarization, respectively,  as shown in Figs. \ref{fig:cp}(a) and (c), which is originated from the distributions of helicity in c.m. frame [see Fig. \ref{fig:pol-xi2}]. In the next scattering of $e'$ and $\gamma_0$, the produced $\gamma''$ with $\xi''_\gamma$ and $e''$ with $\zeta''_e$ can be polarized through both helicities of
$e'$ and $\gamma_0$ via the single or multiplied terms of $\zeta_2$ and $\xi_2$ in $G_2^\gamma,G_2^e$ and $G_3^\gamma, G_3^e$, resulting in the almost completely polarized $\gamma''$ [see Figs. \ref{fig:cp}(a) and (b)]  and the enhancement of $\cos{\theta}$-dependent polarization of $e''$ [see Figs. \ref{fig:cp}(c) and (d)]. The slight energy-dependence of $\xi'_\gamma, \xi''_\gamma$ and $\zeta'_e, \zeta''_e$, as shown in Figs. \ref{fig:cp}(b) and (d), is induced by the non-zero linear polarization from $G_3^\gamma$ which involves a term independent of polarization.
 The collisions between polarized $\gamma'$ or $\gamma''$ and $\gamma_0$ result in the production of BW electron (positron) with highly longitudinal polarization induced by distributed left-hand helicity, as indicated by the sketch of helicity transfer in BW process in Fig. \ref{fig:sketch}. The steep variation of $\zeta'_-$ and $\zeta''_-$ near threshold energy ($\omega =m_e$) originates from the intrinsic helicity transfer of BW process with two right-hand photons \cite{Zhao2023}.

\begin{figure}[!t]
	\setlength{\abovecaptionskip}{0.2cm}
	\centering\includegraphics[width=0.9\linewidth]{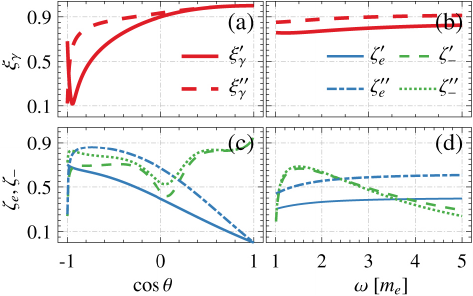}
	\caption{Similar to Fig. \ref{fig:cp} but for $\gamma_0$ with polarization of $\xi^{(0)}_2=\xi^{(0)}_3=0.71$.
}
	\label{fig:clp}
\end{figure}
The polarization scenario of $\gamma_0$ with both degree of linear polarization and circular polarization are also considered; see in Fig. \ref{fig:clp}. In this case, the produced $\gamma'$ with $\xi'_\gamma$ can be polarized via both single $\xi_2$-associated terms in $G_2^\gamma$ and single $\xi_3$-associated term in $G_3^\gamma$, namely, $\gamma'$  is produced with the decrease of circular polarization and meanwhile the increase of linear polarization, leading to the slight diminishment of $\cos{\theta}$-dependent $\xi'_\gamma$; \ref{fig:clp}(a) and (b). While the produced $e'$ with $\zeta'_e$ is mainly polarized through helicity transfer of photon via single $\xi_2$-associated terms in $G_2^e$ since the minor contribution from single $\xi_2$-associated terms in $G_3^e$ [see Fig. \ref{fig:pol-xi2}(d)], leading to the remarkable diminishment of $\zeta'$; see \ref{fig:clp}(c) and (d).  The resulted enhancement of $\xi''$ and $\zeta''$ in this case is similar to the polarization transfer in the case of Fig. \ref{fig:cp}.
Due to the diminished polarization of $\gamma'$ and $\gamma''$ with both circular and linear polarization shown in \ref{fig:clp}(a) and (b), $\zeta'_e$ and $\zeta''_e$ become remarkably $\cos{\theta}$- and $\omega$-dependent; see Figs. \ref{fig:clp}(c) and (d). Therefore, these analytical calculations indicate that the polarization of background  photons  can be transferred throughout the electron-seeded cascade of ICS and BW process.

\section{Polarization transfer in the cascaded processes from MC simulations}\label{simulation}
To numerically model the cascaded processes suggested in Fig. \ref{fig:sketch}, we perform numerical simulations with multi-parameter MC method developed in our previous works \cite{Zhao2022,Lu2022}. This approach permits the incorporation of beam effects of colliding particles, including arbitrary distributions of energy, divergence angle and polarization, and to resolve the final-state angle and polarization of scattered particles from the polarized differential cross sections. The methodology of our MC simulation mainly includes two modules: (1) paring of the colliding particles within the three-dimensional cells using the no-time-count method \cite{Gaudio2020}, and (2) performing the single binary scattering of the paired particles using the MC sampling from the cross sections (see details in the Appendix \ref{appB}). The first module is implemented in the lab frame, while it is in the c.m. frame for  the second module. To this end, the four-momenta of each paired particles in lab frame should be transformed to the ones of the c.m. frame by Lorentz boost along their own c.m. frame velocity $\bm{\beta}_{cm}$. After finishing the scattering of all paired particles at a time step, the second module will return the produced particles with four-momenta of the lab frame. Then the produced particles together with the non-scattered particles are paired and put into scattering within the two modules at the next time step.
In the results of simulation below, $\varepsilon_\gamma$ and $\varepsilon_e$ are denoted as the energy of the photon and the kinetic energy of the electron from ICS, respectively, and $\varepsilon_{pair}$ as the energy of the electron (positron) in the BW process.

We denote the statistical polarization of ICS photons and electrons from MC simulations as $P_\gamma$ and $P_e$, respectively. With the determined energy and momenta of the final-state particles, the statistical polarization can be determined by MC sampling with transition probabilities of polarization given by the differential cross section, as shown in Eq. (\ref{dsig}). Specifically, we obtain
\begin{subequations}\label{polbeam}
\begin{align}
P_\gamma=&\sqrt{(\bar{\xi}'_1)^2+(\bar{\xi}'_2)^2+(\bar{\xi}'_3)^2},\\
P_e=&\sqrt{(\bar{\zeta}'_1)^2+(\bar{\zeta}'_2)^2+(\bar{\zeta}'_3)^2},
\end{align}
\end{subequations}
where $\bar{\xi}'_i$ and $\bar{\zeta}'_i$ are the averaged components of the Stokes parameter and spin components over the particle number. These components are obtained by MC sampling according to the transition probabilities in Eq. (\ref{dsig}). We obtain the statistical polarization of BW electrons (positrons) in a similar way, which we denote as $P_{pair}$.

\begin{figure}[!t]
	\setlength{\abovecaptionskip}{0.2cm}
	\centering\includegraphics[width=0.9\linewidth]{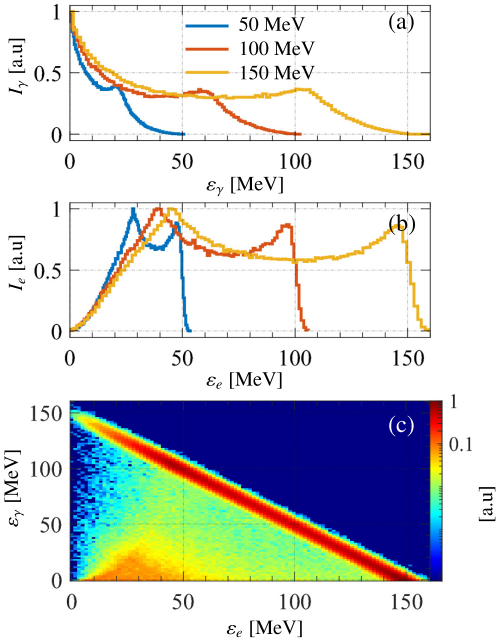}	
	\caption{Normalized energy spectra of scattered photons (a) and electrons (b)  from ICS in laboratory frame, three lines correspond to  seed electrons with peak energy of $E_0=50,100,150$ MeV with energy spread of $5\%$, respectively. (c) Correlated spectra of scattered photons and electrons, corresponding to $E_0=150$ MeV.
The background photons are initialized with $\xi^{(0)}_2=1$, and  IT$\simeq1$ ms is run in the simulations.}
	\label{fig:spectrum}
\end{figure}
In the setups of the electron-seeded cascade between the polarized processes of ICS and BW, we randomize the background soft photons by a spectral distribution of $I_\gamma=I_{\gamma,0}(\varepsilon_\gamma/\varepsilon_{\rm{min}})^{-\alpha}$ with $\varepsilon_{\rm{min}}=0.001$ MeV and a spectral index of $\alpha=1.2$. This mimics the spectral distribution of the M87 black hole \cite{Chen2018} and has an average energy of $\bar{\varepsilon}_{\gamma_0}=0.0361$ MeV. The background soft photons have a uniform density of $\sim1.4\times10^{16}$ cm$^{-3}$. We initialize $10^7$ seed electrons with a spatially uniform density of $\sim1.8\times10^{16}$ cm$^{-3}$, and assume the interaction region to be homogeneous and infinite to neglect the boundary effects of space charge force. We assume a Gaussian spectral distribution of $I_e=I_{e,0}\exp{[-(\varepsilon_e-E_0)/2\sigma_0^2]}$ for the seed electrons. The seed electrons, initialized with root-mean-square divergence angle of 0.1 rad, propagate along $-\hat{z}$ direction and collide with the counter-propagation background photons with the same divergence angle to produce gamma-ray photons via ICS, which later trigger the cascaded BW pair productions by collisions with background photons. The seed electrons are non-polarized, and the background soft photons are initialized with different polarization scenarios to mimic polarized ones produced in jet structures by synchrotron radiation \cite{Nava2016}. This scenario of initial polarization will lead to the production of polarized photons and electrons in ICS, as discussed in Sec. \ref{theory}. Interaction time (denoted as IT) can be tuned to terminate the simulation. The longer duration of $\rm{IT}$  implies the longer series of multiple ICS and the corresponding BW pair production in the cascaded processes.

The energy spectra of scattered photons and electrons from ICS are shown in Fig. \ref{fig:spectrum}, and each spectrum contains more than $10^6$ scattered photons or scattered electrons. Since the total cross section only can be modified by the multiplied initial polarization of photon and electron, namely, $\xi_2\zeta_2$ and $\xi_2\zeta_3$ [see Eq. (\ref{sigtot}) and Fig. \ref{fig:sigtot}], the single polarization of background photons has no modification to the spectra and the later only depend on the initial spectrum of seed electrons. Due to the rapid down ramp of total cross section of Compton scattering near $\omega=0.1m_e$ ,  ICS results in the broaden saddle-shaped spectra of scattered electrons from the initial quasienergetic spectrum of seed electrons, and the double-ramp spectra of scattered photons; see Figs. \ref{fig:spectrum}(a) and (b). In the correlated energy spectrum of $\varepsilon_e$ and $\varepsilon_\gamma$, the linear distribution with energy spread of initial electrons implies the energy conservation from the single scattering of each seed electron, and the diffusion distribution below the linear one originates from the multiple ICS; see Fig. \ref{fig:spectrum}(c). Note that the ratio of scattered particles from multiple ICS to ones from single ICS increases with the interaction time, which will lead to the enhancement of polarization of photons and electrons according to the theoretical prediction in Sec. \ref{theory}.
Note that the energy spectra of Compton electrons (photons) are dominated by the yields of $e'(\gamma')$ in the series of mupltiple ICS [see Fig. \ref{fig:multi-ncs}(a)  in Appendix \ref{appC}].

\begin{figure}[!t]
	\setlength{\abovecaptionskip}{0.2cm}
	\centering\includegraphics[width=0.9\linewidth]{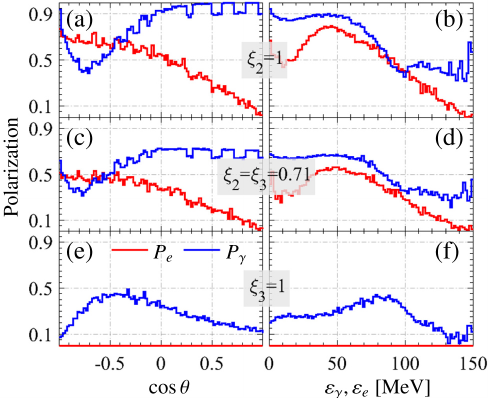}
	\caption{ Polarization curves of first-generation photons and electrons from ICS vs scattering angle and laboratory energy for different polarization scenarios of $\gamma_0$: (a) and (b) $\xi^{(0)}_2=1$; (c) and (d) $\xi^{(0)}_2=\xi^{(0)}_3=0.71$; (e) and (f) $\xi^{(0)}_3=1$.
The seed  electrons are initialized with $E_0=150$ MeV and $5\%$ energy spread.}
	\label{fig:pol-curve}
\end{figure}
The analytical polarization transfer discussed in Sec. \ref{theory} can be retrieved by the statistical polarization from realistic MC simulations in both c.m. frame and laboratory frame.
For polarized $\gamma_0$ with $\xi^{(0)}_2=1$ and $\xi^{(0)}_2=\xi^{(0)}_3=0.71$, polarization curves of $P_e$ and $P_\gamma$ versus $\cos{\theta}$ present the similar behavior to the analytical ones shown in Figs. \ref{fig:cp} and \ref{fig:clp}, except for the different amplitudes due to the modification of polarization variations from the high-order series of multiple ICS (see Fig. \ref{fig:multi-ncs}(b) in Appendix \ref{appC}); see Figs. \ref{fig:pol-curve}(a) and (c). In terms of  laboratory energy $\varepsilon_e$, the scattered electrons are polarized at maximum in the energy region of deep ICS, namely, the energy of seed electrons are broaden downward to the left saddle of the spectrum shown in Fig. \ref{fig:spectrum}(b), after that  $P_e$ decays linearly to zero at the elastic scattering limit where there is no energy transfer; see Fig. \ref{fig:pol-curve}(b) and (d). In comparison, the scattered photons in ICS have the stair-type polarization curves versus energy $\varepsilon_\gamma$, and the polarization of background photons can be more easily transferred to scattered photons even at elastic scattering limit. In terms of polarized $\gamma_0$ with $\xi^{(0)}_3=1$, the polarization transfer still occurs between incident and scattered photons through non-zero single $\xi_3$-associated term in $G_3^\gamma$, leading to the production of photons with linear polarization, while the polarization transfer to scattered electrons is prohibited due to the vanish of helicity transfer; see Fig. \ref{fig:pol-curve}(e) and (f).

\begin{figure}[!t]
	\setlength{\abovecaptionskip}{0.2cm}
	\centering\includegraphics[width=0.9\linewidth]{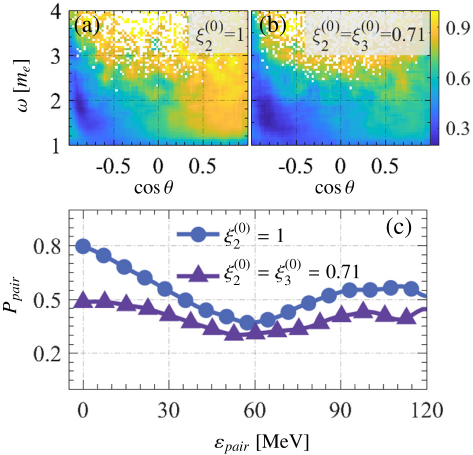}
	\caption{(a) and (b) Polarization distributions of first-generation BW electrons $P_{pair}$ in c.m. frame, produced from ICS between first-generation photons and background photons with polarization scenarios of $\xi^{(0)}_2=1$ and $\xi^{(0)}_2=\xi^{(0)}_3=0.71$. (c) Polarization curves of BW electrons vs kinetic energy $\varepsilon_{pair}$ in laboratory frame. }
	\label{fig:pol-pair}
\end{figure}
The cascaded collisions between polarized first-generation gamma-ray photons and background photons result in the  production of first-generation BW pairs; see Fig. \ref{fig:pol-pair}. According to Eq. (\ref{zpf}), the polarization of BW electron originates from helicity transfer through both helicity of a single photon and cross product of $\bm{\xi}^{(1)}$ and $\bm{\xi}^{(2)}$ of two colliding photons. For polarized $\gamma_0$ with $\xi^{(0)}_2=1$,   collisions between $\gamma_0$ and first-generation gamma-ray photons with circular polarization of right-hand helicities result in the productions of longitudinally polarized BW electrons (positrons), and the asymmetric distribution of $P_{pair}$ results from the $\cos{\theta}$-dependent $\zeta'_\gamma$ and $\zeta''_\gamma$ of first-generation gamma-ray photons; see Figs \ref{fig:pol-pair}(a).
For polarized $\gamma_0$ with $\xi^{(0)}_2=\xi^{(0)}_3=0.71$, collisions between $\gamma_0$ and first-generation gamma-ray photons result in the similar distribution of $P_{pair}$ but the diminished polarization; see Fig. \ref{fig:pol-pair}(b).
Because of the zero contribution of dot product of $\bm{\xi}^{(1)}$ and $\bm{\xi}^{(2)}$ to the final polarization of BW pairs, for polarized $\gamma_0$ with linear polarization of $\xi^{(0)}_3=1$, the cascaded collisions between $\gamma_0$ and first-generation $\gamma$-ray photons with linear polarization  [see Fig. \ref{fig:pol-curve}(f)] produce the non-polarized pairs. The polarization curves of BW electrons versus $\varepsilon_{pair}$ are none-decayed and fluctuant because of the stair-shape ones of first-generation gamma-ray photons and the varying tendency of $P_{pair}$ is caused by the stair-type $P_\gamma$ with corresponding $\varepsilon_\gamma$ above the threshold energy of BW process; Fig. \ref{fig:pol-pair}(c). Since the longitudinal polarization is dominated in the produced first-generation BW electrons, the decrease of $\xi^{(0)}_2$ leads to the decreased $P_{pair}$. It is worth noting that because the polarization is dominated by the first three series of multiple ICS, the variations of $P_\gamma$ and thus the variations of  $P_{pair}$  are almost independent of the duration of $\rm{IT}$ (see Fig. \ref{fig:duration} in Appendix \ref{appC}).

\begin{figure}[!t]
	\setlength{\abovecaptionskip}{0.2cm}
	\centering\includegraphics[width=0.9\linewidth]{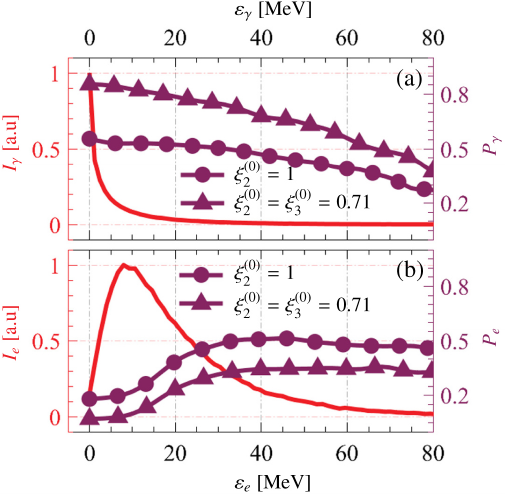}
	\caption{(a) Polarization curves ($P_\gamma$) for two polarization scenarios of $\gamma_0$: $\xi^{(0)}_2=1$ and $\xi^{(0)}_2=\xi^{(0)}_3=0.71$, and spectrum $I_\gamma$ of second-generation photons produced from ICS between BW electrons shown in Fig. \ref{fig:pol-pair} and background photons. (b) Similar to (a) but for second-generation electrons.}
	\label{fig:pol-2nd}
\end{figure}
The produced first-generation BW electrons (positrons) proceed to collide with background photons to produce the second-generation photons and electrons via ICS; see Fig. \ref{fig:pol-2nd}. The second-generation photons are produced with exponential spectrum compared to the double-ramp one of first-generation photons [see Fig. \ref{fig:spectrum}(a)], and the average energy of the second-generation photons is about $\bar{\varepsilon}_\gamma\simeq7.5$ MeV which terminates the second-generation BW process significantly since the c.m. energy $\omega\approx\sqrt{\bar{\varepsilon}_{\gamma_0}\bar{\varepsilon}_\gamma}\simeq m_e$. In comparison to the non-decayed polarization curve of first-generation photons shown in Fig. \ref{fig:pol-curve}, $P_\gamma$ of second-generation photons decay linearly versus $\varepsilon_\gamma$; see Fig. \ref{fig:pol-2nd}(a). Because of the left-hand helicities (negative $\zeta_{-,2}$) of BW electrons, $P_\gamma$ of second-generation photons produced by polarized $\gamma_0$ with $\xi^{(0)}_2=1$ is diminished compared to one produced by polarized $\gamma_0$ with $\xi^{(0)}_2=\xi^{(0)}_3=0.71$. This is because that the polarization transfer from negative helicities of BW electrons and positive helicity of $\gamma_0$ leads to the cancelation each other according to Eq. (\ref{Ggam}), while for partially linear polarization of $\gamma_0$, the $\xi_3$-associated term in Eq. (\ref{Ggam}) leads to the production of highly-polarized second-generation photons.
The spectrum of second-generation electrons are peaked due to the typical spectrum of BW pairs; see Fig. \ref{fig:pol-2nd}(b).
The polarization curves $P_e$ of second-generation electrons are saturated at high-energy region compared to the decayed one of first-generation electrons. $P_e$ of second-generation electrons produced by $\gamma_0$ with $\xi^{(0)}_2=1$ is higher than the one from $\xi^{(0)}_2=\xi^{(0)}_3=0.71$ due to the helicity-dominated polarization transfer for electrons. Therefore, in terms of produced photons from ICS, the saturated polarization curves in first-generation cascade can be transformed to the linearly decayed one in second-generation cascade, and it is the inverse case for produced electrons from ICS.

\section{CONCLUSION}\label{summary}
In summary, we investigate the polarization transfer in the cascaded processes of ICS and BW pair production through theoretical analysis and numerical simulations. Analytical analysis show that in the cascade of multiple ICS and BW process, the polarization of background soft photons can be effectively transferred to the final-state particles of the first-generation cascade owing to helicity transfer. With multi-parameter MC numerical simulations, we clarify the energy-dependent polarization curves in the electron-seeded cascade of ICS and BW process in detail. The circular polarization of background photons is transferred to gamma-ray photons via multiple ICS, leading to the production of first-generation gamma-ray photons with non-decayed polarization curves and first-generation electrons with linearly decayed ones. The polarized gamma-ray photons trigger the production of polarized BW pairs with fluctuant polarization curves, and the degree of polarization depends merely on the circular polarization of background photons. The first-generation BW electrons (positrons) trigger the second-generation cascade via ICS, which produces distinct polarization curves of photons and electrons compared to those in the first-generation cascade. The energy loss of second-generation photons leads to the termination of the cascade, thus precluding the production of second-generation BW pairs. This understanding of the polarized cascade between ICS and BW process is beneficial for investigating high-intensity laser-driven QED plasmas and related research on high-energy astrophysical phenomena such as gamma-ray bursts and active galactic nuclei.

\section{ACKNOWLEDGEMENT}
The work is supported by the National Natural Science Foundation of China (Grants No. 12022506, No. U2267204, 12275209, 12105217), the Foundation of Science and Technology on Plasma Physics Laboratory (No. JCKYS2021212008), the Open Foundation of Key Laboratory of High Power Laser and Physics, Chinese Academy of Sciences (SGKF202101), and the Shaanxi Fundamental Science Research Project for Mathematics and Physics (Grant No. 22JSY014).
\appendix
\section{Completely polarized cross section of Compton scattering} \label{appA}
The polarized cross section of Compton scattering is formulated in the center-of-momentum (c.m.) frame as (relativistic units with $\hbar=c=1$ are used throughout) \cite{Berestetskii1982}
\begin{eqnarray}
\frac{d\sigma_{\rm{c}}}{dtd\phi}=\frac{r_e^2}{32\pi^2}\frac{|M_{fi}|^2}{(s-1)^2},\label{dsigCompton}
\end{eqnarray}
where $r_e$ is classical electron radius, $|M_{fi}|^2$ is scattering amplitude expressed by photon and electron density matrixes. $s$ and $t$ are normalized Mandelstam invariants defined as
\begin{subequations}\label{stu}
\begin{eqnarray}
s&=&(k_1+p_1)^2/m_e^2=(k_2+p_2)^2/m_e^2,\\
t&=&(p_2-p_1)^2/m_e^2=(k_2-k_1)^2/m_e^2,\\
u&=&(p_1-k_2)^2/m_e^2=(p_2-k_1)^2/m_e^2.
\end{eqnarray}
\end{subequations}
In c. m. frame, $s=(\omega+\epsilon)^2/m_e^2$ with $\omega$ of photon energy and $\epsilon=\sqrt{\omega^2+m_e^2}$ of electron energy. The density matrixes can be expanded on the defined unit 4-vectors base, and by calculating the trace of matrix product, $|M_{fi}|^2$ can be expressed explicitly with Stokes parameters $\bm{\xi}$ and $\bm{\xi}'$ of initial and final photons, and mean spin vectors $\bm{\zeta}$ and $\bm{\zeta}'$  of initial and final electrons.

Using the denotations of $x=s-1$ and $y=1-u$, the differential cross section of Compton scattering can be expressed as
\begin{align}\label{dsig}
  \frac{d\sigma_{\rm{c}}}{dyd\phi} = \frac{r_e^2}{4x^2}\left[F_{\rm{c}}+\sum_{i=1}^{3}(G_i^\gamma\xi'_i+G_i^e\zeta'_i)+\sum_{i,j=1}^{3}H_{i,j}\zeta'_i\xi'_j\right]
\end{align}
where $\zeta'_i$ and $\xi'_i$ are components of $\bm{\zeta}'$ and $\bm{\xi}'$.
After integration over $y$ (between $x/(x+1)$ and $x$) and $\phi$, one obtains the total cross section of Compton scattering
\begin{eqnarray}\label{sigtot}
  \sigma_{\rm{c}} &=& \frac{\pi r_e^2}{4x^2}\left(\sigma_0+\sigma_1\right)
\end{eqnarray}
with
\begin{subequations}
	\begin{align}
		\sigma_0=&\frac{8(x^2-4x-8)\log{(x+1)}}{x}+\frac{4(x^3+18x^2+32x+16)}{(x+1)^2}\\
\sigma_1=&\left[4(x + 1)^2(x + 2)\log{(x + 1)} - 2x(5x^2 + 8x + 4)\right]\xi_2\zeta_2\nonumber\\
+&\left[(3x^2 + 12x+ 8)\sqrt{x + 1}-8x^2-16x-8\right]\xi_2\zeta_3
	\end{align}
\end{subequations}
\begin{figure}[!t]
	\setlength{\abovecaptionskip}{0.2cm}
	\centering\includegraphics[width=0.9\linewidth]{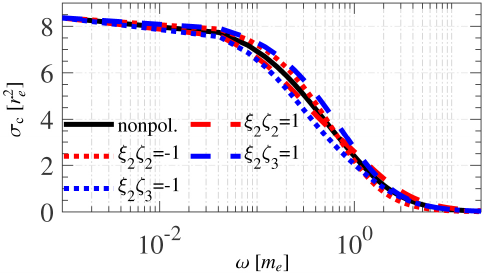}
	\caption{The total cross section of Compton scattering, calculated from Eq. (\ref{sigtot}),  corresponding to different initial polarization of photon and electron. }
	\label{fig:sigtot}
\end{figure}
Since $\sigma_{\rm{c}}$ depends on the products of $\xi_2\zeta_2$ and $\xi_2\zeta_3$ with circular polarization ($\xi_2$) of photon, and both the longitudinal-spin polarization ($\zeta_2$) and transverse-spin polarization ($\zeta_3$) of electron, the non-polarized $\sigma_{\rm{c}}$ can be decomposed into four scenarios, as shown in  Fig. \ref{fig:sigtot}. See that the initial polarization slightly modify the amplitude of $\sigma_{\rm{c}}$ at the down ramp.

By summation over final electron spin, the differential cross section is
\begin{eqnarray}\label{dsigsum}
\frac{d\bar{\sigma}_{\rm{c}}}{dyd\phi}=\frac{r_e^2}{4x^4y^2}F_{\rm{c}},
\end{eqnarray}
where $F_{\rm{c}}=\left(F_0+F_1\xi_1+F_{22}\xi_2\zeta_2+F_{23}\xi_2\zeta_3+F_3\xi_3\right)$ with
\begin{subequations}
	\begin{align}
F_0=&4 \left(x^3 y-4 x^2 y+4 x^2+x y^3+4 x y^2-8 x y+4 y^2\right)\\
F_1=&16 (x-y) (xy-x+y) \sin{(2\phi)}\\
F_{22}=&4 (x-y) \left(x^2y-2x^2+x y^2+2 y^2\right)\\
F_{23}=&8y(x-y)\sqrt{x-y}\sqrt{x y-x+y}\\
F_3=&16\left(x^2 y-x^2-x y^2+2 x y-y^2\right)\cos{(2\phi)}
	\end{align}
\end{subequations}

The concrete expressions of $G_i^\gamma$ in Eq. (\ref{dsig}) are
\begin{subequations}\label{Ggam}
\begin{align}
G_1^\gamma&=-8\zeta_1\xi_2\frac{x^3y-x^3-2x^2y^2+3x^2y+xy^3-3xy^2+y^3}{xy^2\sqrt{(x-y)(xy-x+y)}} \nonumber\\
       &+8\frac{\xi_1}{xy^2}(xy^2-2xy+2y^2),\\
G_2^\gamma&=8\zeta_1\xi_1\sqrt{xy-x+y}\frac{x^3y-2x^2y^2+xy^3}{x^3y^2\sqrt{x-y}}\nonumber\\
    +&8\zeta_2\xi_3\frac{x^3y-x3-2x^2y^2+3x^2y+xy^3-3xy^2+y^3}{x^3y^2}\nonumber\\
       +&4\zeta_2\frac{x^4y-4x^3y+4x3-x^2y^3+8x^2y^2-12x^2y-4xy^3+12xy^2-4y^3}{x^3y^2}\nonumber\\
       +&4\zeta_3\xi_3\frac{\sqrt{(x-y)(xy-x+y)}(x+2)(x-y)^2}{x^3y^2}\nonumber\\
       +&8\zeta_3\frac{\sqrt{(x-y)(xy-x+y)}(x-y)(xy-2x+2y)}{x^3y^2}\nonumber\\
       +&4\xi_2\frac{x^4y-2x^4+2x^3y+x^2y^3-2x^2y^2+2xy^3}{x^3y^2},\\
 G_3^\gamma&=8\zeta_2\xi_2\frac{x^3y-x3-2x^2y^2+3x^2y+xy^3-3xy^2+y3}{x^3y^2}\nonumber\\
       +&4\zeta_3\xi_2\frac{\sqrt{(x-y)(xy-x+y)}(x+2)(x-y)^2}{x^3y^2}\nonumber\\
       +&8\xi_3\frac{x^3y^2-2x^3y+2x3+2x^2y^2-4x^2y+2xy^2}{x^3y^2}\nonumber\\
       +&16\frac{x^3y-x3-x^2y^2+2x^2y-xy^2}{x^3y^2},
\end{align}
\end{subequations}
and the concrete expressions of $G_i^e$ in Eq. (\ref{dsig}) are
\begin{widetext}
\begin{subequations}\label{Gele}
\begin{align}
G_1^e&=16\zeta_1\xi_3\frac{x^2y-x^2-xy^2+2xy-y^2}{x^2y^2}\nonumber\\
  &+8\zeta_1\frac{x^2y^2-2x^2y+2x^2+2xy^2-4xy+2y^2}{x^2y^2}\nonumber\\
       &+8\zeta_2\xi_1\sqrt{x-y}\frac{x^3y-x^3-x^2y^2+2x^2y-xy^2}{x^2y^2\sqrt{xy-x+y}}\\
G_2^e&=-8\zeta_1\xi_1\sqrt{x-y}\frac{x^4y^2-3x^4y+2x^4-x^3y^3+6x^3y^2-6x^3y-3x^2y^3+6x^2y^2-2xy^3}{x^4y^2\sqrt{xy-x+y}}\nonumber\\
       &+16\zeta_2\xi_3\frac{x^3y^2-3x^3y+2x^3-x^2y^3+6x^2y^2-6x^2y-3xy^3+6xy^2-2y^3}{x^4y^2}\nonumber\\
       &+4\zeta_2\frac{x^5y-2x^5+2x^4y+x^3y^3-6x^3y^2+12x^3y-8x^3+6x^2y^3-24x^2y^2+24x^2y+12xy^3-24xy^2+8y^3}{x^4y^2} \nonumber\\
       &+8\zeta_3\xi_3\sqrt{x-y}\sqrt{xy-x+y}\frac{x^3y-x^2y^2+4x^2y-4x^2-4xy^2+8xy-4y^2}{x^4y^2}\nonumber\\
       &8\zeta_3\sqrt{x-y}\sqrt{xy-x+y}\frac{x^3y+x^2y^2-4x^2y+4x^2+4xy^2-8xy+4y^2}{x^4y^2}\nonumber\\
       &+4\xi_2\frac{x^5y-4x^4y+4x^4-x^3y^3+8x^3y^2-12x^3y-4x^2y^3+12x^2y^2-4xy^3}{x^4y^2}\\
G_3^e&=16\zeta_1\xi_1\frac{x^4y-x^4-2x^3y^2+3x^3y+x^2y^3-3x^2y^2+xy^3}{x^4y^2}\nonumber\\
       &-8\zeta_2\xi_3\sqrt{x-y}\frac{x^5y-x^5-x^4y^2+2x^4y+3x^3y^2-8x^3y+4x^3-4x^2y^3+16x^2y^2-12x^2y-8xy^3+12xy^2-4y^3}{x^4y^2\sqrt{xy-x+y}}\nonumber\\
       &-8\zeta_2\sqrt{x-y}\frac{x^4y^2-x^4y+x^3y^3-4x^3y^2+8x^3y-4x^3+5x^2y^3-16x^2y^2+12x^2y+8xy^3-12xy^2+4y^3}{x^4y^2\sqrt{xy-x+y}}\nonumber\\
       &+16\zeta_3\xi_3\frac{x^3y^2-3x^3y+2x^3-x^2y^3+6x^2y^2-6x^2y-3xy^3+6xy^2-2y^3}{x^4y^2}/\nonumber\\
       &+8\zeta_3\frac{x^4y^2-2x^4y+6x^3y-4x^3+2x^2y^3-12x^2y^2+12x^2y+6xy^3-12xy^2+4y^3}{x^4y^2}\nonumber\\
       &-8\xi_2\sqrt{x-y}\frac{x^4y^2-3x^4y+2x^4-x^3y^3+6x^3y^2-6x^3y-3x^2y^3+6x^2y^2-2xy^3}{x^4y^2\sqrt{xy-x+y}}
\end{align}
\end{subequations}
\end{widetext}
In c.m. frame, the normalized invariants $x$ and $y$ are expressed as $x=2\omega(\omega+\epsilon)/m_e^2$ and $y=2\omega(\epsilon+\omega\cos{\theta})/m_e^2$ with scattering angle of $\cos{\theta}$ between initial and scattered photons.

\begin{figure}[!t]
	\setlength{\abovecaptionskip}{0.2cm}
	\centering\includegraphics[width=0.9\linewidth]{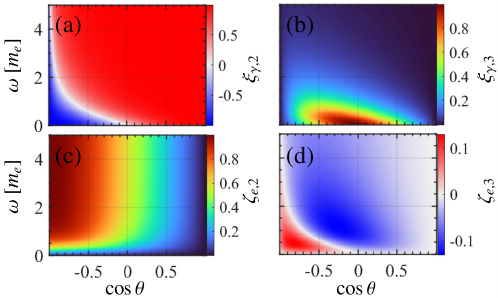}
	\caption{(a) Stokes parameters of final Compton photon. (b) Spin components of final Compton electron. The results are obtained by Eqs. (\ref{xif}) and (\ref{zef}) with $\xi^{(0)}_2=1$ and $\zeta_i=0$.}
	\label{fig:pol-xi2}
\end{figure}
The polarization states of scattered photon and electron in Compton scattering can be clarified through their mean polarization vectors analytically, as shown in Fig. \ref{fig:pol-xi2}. See that in the Compton scattering with right-hand photon ($\xi^{(0)}_2=1$) and non-polarized electron, the scattered  electron can be polarized with right-hand helicities in angle region of $\cos{\theta}<0$, and the photon polarization reverses its circular polarization from right-hand to left-hand [see Figs. \ref{fig:pol-xi2}(a) and (c)]. The linearly polarized photon can be produced for transition from $\cos{\theta}>0$ to $\cos{\theta}<0$ at low c.m. energy [see Fig. \ref{fig:pol-xi2}(b)], and electron reverses its orientation of transverse polarization while the photon polarization reverses its orientation of circular polarization [see Fig. \ref{fig:pol-xi2}(d)].
The variation of distributed polarization with $\omega$ and $\cos{\theta}$ of c.m. frame indicate that the intrinsic spin angular momentum is not conserved in Compton scattering, which is consistent with the conclusion in  \cite{Ahrens2017}.

For convenience of figuring out the production of polarized BW pairs in cascaded processes, spin components of final-state electrons $\zeta_{-,i}=G^-_{i}/F_{\rm{bw}}$ are presented here
\begin{subequations}\label{Gpair}
\begin{align}
	G^-_{1} &=-\frac{8(v+w)\sqrt{v^2w-v^2+vw^2-2vw-w^2}}{v^2w^2}\nonumber\\
\times&\left(\xi_1^{(1)}\xi_2^{(2)}v-\xi_1^{(2)}\xi_2^{(1)}w\right),\\
    G^-_{2} &=\frac{4 (v+w) \sqrt{v+w}}{v^2 w^2\sqrt{v+w-4}}\left[(-4 v w+4 v+4 w)\xi_2^{(1)}\xi_3^{(2)}\right.\nonumber\\
+&(v^2 w-v w^2+2 v w-4 v+2 w^2-4 w) \xi_2^{(1)} \nonumber\\
+&(4 v w-4 v-4 w) \xi_2^{(2)} \xi_3^{(1)}\nonumber\\
+&\left.(v^2 w-2 v^2-v w^2-2 v w+4 v+4 w)\xi_2^{(2)}\right],\\
G^-_{3}&=v^2w^2\sqrt{x + y - 4}\sqrt{x y - x - y}\nonumber\\
\times&\left[-8 \left(v^3 w^2-3 v^3 w+2 v^3+2 v^2 w^3-7 v^2 w^2+6 v^2y w+v w^4\right.\right.\nonumber\\
-&\left.5 v w^3+6 v w^2-w^4+2 w^3\right)\xi_2^{(1)}\xi_3^{(2)}+16 \left(v^3 w^2-2 v^3 w+v^3\right.\nonumber\\
 +&\left.v^2 w^3-4 v^2 w^2+3 v^2 w-2 v w^3+3 v w^2+w^3\right)\xi_2^{(1)}\nonumber\\
-&8\left(v^4 w-v^4+2 v^3 w^2-5 v^3 w+2 v^3+v^2 w^3-7 v^2 w^2 \right.\nonumber\\
+&\left.6 v^2 w-3 v w^3+6 v w^2+2 w^3\right)\xi_2^{(2)}\xi_3^{(1)}\nonumber\\
+&\left.16 \left(v^3 w^2-2 v^3 w+v^3+v^2 w^3-4 v^2 w^2+3 v^2 w\right.\right.\nonumber\\
-&\left.\left.2 v w^3+3 v w^2+w^3\right)\xi_2^{(2)}\right],
\end{align}
\end{subequations}
here, $(\xi_1^{(1)},\xi_2^{(1)},\xi_3^{(1)})$ and $(\xi_1^{(2)},\xi_2^{(2)},\xi_3^{(2)})$ represent the Stokes parameters of the two colliding photons denoted as '(1)' and '(2)' respectively. In c. m. frame, the Lorentz invariants $v$ and $w$ are expressed as $v=2\omega^2/m_e^2-2\omega\sqrt{\omega^2-m_e^2}\cos{\theta}/m_e^2$ and $w=2\omega^2/m_e^2+2\omega\sqrt{\omega^2-m_e^2}\cos{\theta}/m_e^2$, with $\omega$ the c.m. energy of initial photon and $\cos{\theta}$ the scattering angle between initial photon and final electron.

\section{Construction of the Compton scattering by MC sampling} \label{appB}
In the simulation, the realistic photons and electrons (not macroparticles) are randomized. Considering a collision cell containing $N_\gamma$ photons and $N_e$ electrons, the maximum probability of any particles to collide within a time step $\Delta t$ and an cell volume $\Delta V$ is
\begin{equation}
  P_{max}=2\sigma_Tc\Delta t/\Delta V,
\end{equation}
where $\sigma_T$ is Thomson cross section.
At a time step, the maximum number of photons in a cell that are probable to scatter is $N_{max}=P_{max}N_\gamma N_e$.

After randomly sorting the photons and electrons, respectively, in a cell, the paired photon and electron are selected from the first $N_{max}$ particles in the sorted photon lists. The event probabilities are
\begin{equation}
  P^{i,j}=(\bar{\sigma}_{\rm{c}}c\Delta t/\Delta V)/P_{max}.
\end{equation}
Paired photon and electron are admitted to the Compton scattering based on the acceptance-rejection method, i.e., for a random number $R_0$, the Compton scattering occurs if $P^{i,j}>R_0$.

Assuming the scattering of a photon and an electron with four-momenta $k_1=(\omega_1,\bm{k}_1)_{lab}$ and $p_1=(\varepsilon_1,\bm{p}_1)_{lab}$ in lab frame, respectively. By the Lorentz boost along $\bm{\beta}_{cm}=(\bm{k}_1+\bm{p}_1)/(\omega_1+\varepsilon_1)$, the corresponding four-momenta $k=(\omega,\bm{k})_{cm}$ and $p_e=(\varepsilon,-\bm{k})_{cm}$ in c.m. frame can be obtained, where $\varepsilon=\sqrt{\omega^2+m_e^2}$. Substituting the c.m. energy and polarization into the total cross section $\sigma_{c}$, the event of Compton scattering can be randomly sampled by using the acceptance-rejection method.
\begin{figure}[!t]
	\setlength{\abovecaptionskip}{0.2 cm}
	\centering\includegraphics[width=0.8\linewidth]{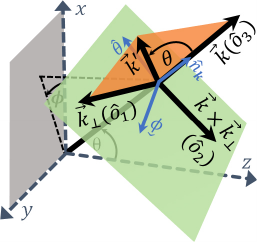}
	\caption{Coordinate system of the Compton scattering in the c.m. frame. $\vec{k}$ and $\vec{k'}$ are the momenta of initial and final-state photon, respectively. $(\hat{n},\hat{\theta},\hat{\phi})$ compose the base of spherical coordinates of $\vec{k}$. ($\hat{o}_1,\hat{o}_2,\hat{o}_3$) compose the base for $\vec{k'}$. $\vec{k}_{\perp}$ is perpendicular to $\vec{k}$ and onto the plane of $\hat{o}_1$ and $\hat{o}_2$. $\vec{k'}$ is the momentum of scattered photon and onto the plane of $\hat{o}_1$ and $\hat{o}_3$.}
	\label{fig:mom}
\end{figure}

As an event of Compton scattering occurs, the c.m. three-momenta $\bm{k}'$ of scattered photon can be determined by the defined coordinate system; see Fig. \ref{fig:mom}. The amplitude of $\bm{k}'$ can be determined by the Madelstam invariant $s_m=(k_1+p_1)^2=m_e^2+2\omega(\omega+\varepsilon)$, resulting in $\omega=|\bm{k}|=(s_m-m_e^2)/2\sqrt{s_m}$.
Define $\hat{\bm{n}}'$ as direction of $\bm{k}'$, which can be determined via the relations of vectors  in the defined coordinate system
\begin{subequations}\label{npe}
\begin{eqnarray}
 \hat{\bm{n}}_\perp&=&\cos\phi\hat{\theta}+\sin\phi\hat{\phi},\\
 \hat{\bm{n}}'&=&\cos{\theta}\hat{n}_k+\sin{\theta}\hat{n}_\perp.
\end{eqnarray}
\end{subequations}
The values of $\theta$ and $\phi$ are determined from differential cross section Eq. (\ref{dsig}) by using random sampling, specifically, by solving the following equation
\begin{subequations}\label{thephi}
\begin{align}
\int_{x/(x+1)}^{y}d\bar{\sigma}_1=&\sigma_{\rm{c}}|R_1|,\\
\int_{0}^{\phi}d\bar{\sigma}_2=&\sigma_{\rm{c}}R_2,
\end{align}
\end{subequations}
where $R_1$ and $R_2$ are uniform random numbers between -1 and 1, and 0 and 1, respectively, and $\cos{\theta}=R_1(y/2\omega^2-\varepsilon/\omega)$. Here, $d\bar{\sigma}_1$ and $d\bar{\sigma}_2$ are defined as integrals of $d\bar{\sigma}_{\rm{c}}^2/dyd\phi$ with respect to $\phi$ and $y$, respectively.

With the determined $\theta$ and $\phi$, the direction of three-momenta $\bm{k}'$ are obtained by Eq. (\ref{npe}). Finally, the energy and momenta of final-state photon (electron) in the laboratory frame are obtained by inverse Lorentz boost
\begin{align}\label{elab}
\omega_{2}=&\omega\gamma_{cm}(1+\beta_{cm}\cos{\theta_u}),\\
\bm{k}_{2}=&\bm{k}'+\frac{\gamma_{cm}^2}{(\gamma_{cm}+1)}(\bm{k}'\cdot\bm{\beta}_{cm})\bm{\beta}_{cm}+\gamma_{cm}\omega\bm{\beta}_{cm},
\end{align}
where $\gamma_{cm}=1/\sqrt{1-\beta_{cm}^2}$ and $\theta_u$ is the angle between $\bm{\beta}_{cm}$ and $\bm{k}'$. The four-momenta $k_2=(\omega_2,\bm{k}_2)_{lab}$ and $p_2=(\varepsilon_2,\bm{p}_2)_{lab}$ are used to construct the subsequent Compton scattering or BW pair production in the similar way illustrated above.

\section{Effects of multiple Compton scattering} \label{appC}
\begin{figure}[!t]
	\setlength{\abovecaptionskip}{0.2cm}
	\centering\includegraphics[width=0.9\linewidth]{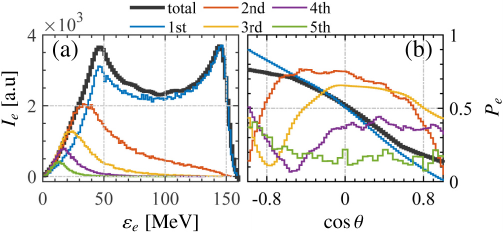}	
	\caption{(a) Spectra of Compton electrons at different series of multiple Compton scattering. (b) Polarization versus scattering angle $\cos{\theta}$ corresponding to Compton electrons of (a). The thick solid lines plot the spectra and polarization of total Compton electrons.}
	\label{fig:multi-ncs}
\end{figure}
The energy spectra of Compton electrons produced at every series of multiple ICS with absolute number are shown in figure \ref{fig:multi-ncs}(a). One can see that although the energy spectra of Compton electrons are dominated by the first-order (denoted as ``1st'' in figure \ref{fig:multi-ncs}(a)) series, the spectra of higher-order series with single peak are non-negligible and modify the low-energy region of total spectra. Apparently, the single-peak spectra of higher-order series shift toward low energy due to the continuous energy loss to Compton photons. Because of the non-negligible yields of higher-order series, the polarization from $e'', e''', \ldots$ can contribute to the total polarization.
One can see that the variations of total polarization along scattering angle can be modified by the higher-order series of Compton scattering around colliding axis ($\cos{\theta}=\pm1$); see figure \ref{fig:multi-ncs}(b).

\begin{figure}[!t]
	\setlength{\abovecaptionskip}{0.2cm}
	\centering\includegraphics[width=0.9\linewidth]{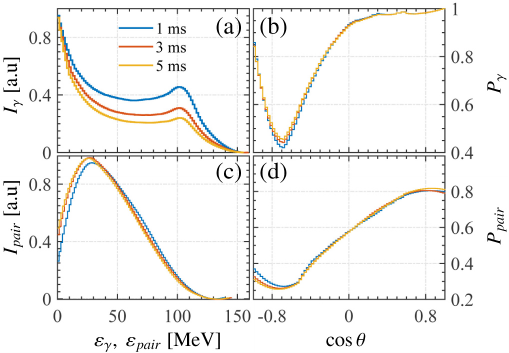}	
	\caption{
(a) and (c) Spectra of first-generation Compton photons and first-generation BW pairs produced by different durations of IT, respectively. (b) and (d) Polarization variations versus scattering angle $\cos{\theta}$ corresponding to electrons in (a) and pairs in (c), respectively.}
	\label{fig:duration}
\end{figure}
In general, as the parameters of the distributed background photons and seed electrons are fixed in a simulation, the longer duration of $\rm{IT}$  implies the longer series of multiple ICS, leading to the larger yields of Compton photons (electrons) and BW pair in the cascaded processes. As the duration of $\rm{IT}$ increases from 1 ms to 5 ms, the normalized spectra $I_\gamma$ of first-generation Compton photons are shifted downward due to the energy loss by the longer series of multiple ICS, as shown in Fig. \ref{fig:duration}(a).  However,  the normalized spectra $I_{pair}$ of first-generation BW pairs have almost negligible variations because of the non-modified profiles of $I_\gamma$; see Fig. \ref{fig:duration}(c). Moreover, because the polarization is dominated by the first three series of multiple ICS, the variations of $P_\gamma$ and thus the variations of  $P_{pair}$  are almost independent of the duration of  $\rm{IT}$; see Figs. \ref{fig:duration}(b) and (d).

\bibliography{refs-cascade}

\end{document}